\documentclass[%
 reprint,
 amsmath,amssymb,
 aps,
]{revtex4-2}

\usepackage{graphicx}
\usepackage{dcolumn}
\usepackage{bm}
\usepackage{siunitx}
\usepackage{pifont} 
\usepackage{booktabs} 

\begin{document}

\preprint{APS/123-QED}

\title{Particle-in-Cell Simulations of Burning ICF Capsule Implosions}

\author{Johannes J. van de Wetering}
\email{vandeweterin1@llnl.gov}
\author{Justin R. Angus}
\author{W. Farmer}
\author{V. Geyko}
\author{D. Ghosh}
\author{D. Grote}
\author{C. Weber}
\author{G. Zimmerman}

\affiliation{Lawrence Livermore National Laboratory, Livermore, CA 94551, USA}
\date{\today}


\begin{abstract}
Anomalies observed in the neutron spectral shift of high-yield shots at the National Ignition Facility (NIF) suggest the presence of suprathermal ions \cite{hartouni_evidence_2023}, implying that kinetic effects play a significant role in burning inertial confinement fusion (ICF) plasmas. Furthermore, recent measurements of reaction-in-flight (RIF) neutrons offer a direct probe of the stopping power in the burning fuel region of high energy alpha particles and up-scattered fuel ions. We have developed the particle-in-cell code PICNIC, an \textit{exactly} energy-conserving particle-in-cell Monte-Carlo collision (PIC-MCC) code to simulate the burn stage in ICF. We present results from 1D spherical simulations of NIF shot N210808. We find that the suprathermal ions generated by large-angle Rutherford and nuclear elastic scattering (NES) with fusion alphas produce an alpha knock-on neutron (AKN) signal consistent with experiments. We also find that the inclusion of large-angle scattering physics does not explain the anomalously large spectral shift observed in experiment.
\end{abstract}


\maketitle


\section{Introduction}

Ever since the first ignition result at the National Ignition Facility (NIF) \cite{zylstra_burning_2022, zylstra_experimental_2022}, the NIF has provided access to the burning plasma regime, opening new research avenues to further optimize yield. For instance, recent measurements of the neutron spectra of burning shots have exhibited anomalously large spectral shifts in the D--T primary spectrum beyond what is expected from a purely thermonuclear plasma \cite{hartouni_evidence_2023}. This observation suggests the presence of a significant population of suprathermal fuel ions and highlights the importance of understanding kinetic effects in this regime. Furthermore, in burning D/T plasmas it is known that large-angle elastic scattering of alpha particles against thermal D/T fuel ions generates suprathermal ion populations, leading to significant modifications to the emitted neutron spectrum and can influence overall yield \cite{ryutov_energetic_1992,helander_formation_1993,fisher_fast_1994,ballabio_-particle_1997,kallne_observation_2000,korotkov_observation_2000,matsuura_effect_2006,matsuura_distortion_2007,zaitsev_suprathermal_2007}. A spectral feature of particular interest is the high energy reaction-in-flight (RIF) alpha knock-on neutron (AKN) tail in the 15.5--18 MeV range, which has been proposed to serve as a direct probe of hotspot conditions and stopping powers of alphas and suprathermal fuel ions in the partially degenerate burning plasmas in ICF implosions \cite{hayes_reaction--flight_2015,jeet_diagnosing_2024,gatu_johnson_learning_2024}.

The particle-in-cell (PIC) code PICNIC is used in this work to investigate the role of kinetic effects in burning ICF capsule implosions \cite{angus_implicit_2023,angus_implicit_2024}. In this letter we first give an overview of the physics modules included in PICNIC used in this work. We then provide a brief discussion of neutron spectra in ICF and the components we include in our study. Finally, we present 1D spherical simulation results of the full burn stage of the first NIF ignition shot N210808.

\section{Methods}

PICNIC is a fully implicit, \textit{exactly} energy-conserving, electromagnetic and relativistic PIC-Monte-Carlo collision (PIC-MCC) code that supports planar, 1D/2D cylindrical and 1D spherical geometries \cite{angus_implicit_2024}. For all scattering routines, PICNIC uses a moment-preserving MCC method that maintains the correct scattering physics between particles of different weights while preserving local momentum and energy conservation \cite{angus_moment-preserving_2025}. This method is especially necessary for problems in spherical geometry with non-local transport (e.g. alpha heating) where macroparticle weights can strongly vary within a cell.

\subsection{Coulomb Scattering}

The Coulomb collision module includes both cumulative small-angle and single large-angle Rutherford scattering for moderately coupled plasmas \cite{angus_binary_2025,bobylev_monte_2013}, which builds upon the work by Turrell \cite{turrell_self-consistent_2015} and uses the quantum mechanical impact parameter corrections derived in \cite{lindhard_relativistic_1996}. The inclusion of large-angle Rutherford scattering physics is necessary to yield the Li-Petrasso $\sim1/\ln\Lambda$ corrections to the Fokker-Planck collision operator \cite{li_fokker-planck_1993} in the moderately coupled plasmas $2\lesssim\ln\Lambda\lesssim10$ relevant to ICF implosions \cite{lindl_review_2014,reichelt_effects_2024}. Single large-angle Rutherford scattering also contributes to the generation of suprathermal ions in burning plasmas, though it is superseded by nuclear elastic scattering (NES) in the relevant multi-MeV energy range.

\subsection{Radiation}

For radiative transport, we have implemented a 3D ray-tracing bremsstrahlung and inverse bremsstrahlung model, building upon on the work by Lavell \cite{lavell_kinetic_2024} which uses the Seltzer-Berger cross section tables \cite{seltzer_bremsstrahlung_1986}. Our implementation differs in a few key ways: For bremsstrahlung emission, we also include the (weighted) ion recoil to simultaneously conserve energy and momentum, rather than only considering the electron as in \cite{lavell_kinetic_2024} which conserves energy but not momentum. For inverse bremsstrahlung we use another simultaneously energy- and momentum-conserving method which distributes the absorbed photon energy and momentum amongst all the electrons in a cell, similar to the implementation of the moment-preserving collision method for weighted particles \cite{angus_moment-preserving_2025}. 

In addition, we use a cutoff frequency set by the plasma frequency $\omega_\text{p}$ in the cell, and we apply a dynamic group velocity of the photon macroparticles set by $v_\text{g}/c = (1-\omega_\text{p}^2/\omega^2)^{1/2}$ for more accurate radiation transport. For simplicity, we do not consider refraction of the rays. 

\subsection{Nuclear Elastic Scattering}

NES refers to the strong force mediated interaction X(Y,Y')X', where X' denotes the recoil of the nucleus X against Y, which both remain in the ground state. Specifically for the single large-angle collisions relevant to multi-MeV suprathermal ion production, NES generally has a cross section roughly one to two orders of magnitude larger than that of Rutherford scattering \cite{helander_formation_1993} and is therefore important to consider for burning D/T plasmas. To include the effects of NES, we follow the approach outlined in \cite{ballabio_-particle_1997}, where the total differential cross section is written as a sum of the pure Rutherford cross section and a nuclear interference term
\begin{align}
	\frac{\text{d}\sigma}{\text{d}\Omega} = \frac{\text{d}\sigma_\text{C}}{\text{d}\Omega} + \frac{\text{d}\sigma_\text{NI}}{\text{d}\Omega}\,.
\end{align}
The Rutherford cross section always dominates at shallow angles, which is already handled by the Coulomb collision algorithm. A minimum cutoff angle of $\theta_\text{min} = \max(\theta_0,20^\circ)$ is chosen as per the procedure in \cite{ballabio_-particle_1997}, where $\theta_0$ is the largest root of the nuclear interference term, to ensure positivity of the nuclear interference term while maintaining sufficient accuracy for the large-angle collisions relevant to suprathermal ion generation.

To model AKN, we have implemented NES for $\alpha$--D and $\alpha$--T using the Okhrimovskyy method for anisotropic elastic scattering \cite{okhrimovskyy_electron_2002} by calculating the total and transport cross sections of the nuclear interference term using the differential cross section tables from the DRESS code \cite{eriksson_calculating_2016}. These tables are based on the optical model for nuclear interference scattering in \cite{ballabio_-particle_1997}.  On top of the scattering with alphas, we also consider NES for D--T following the same procedure using differential cross sections from the ENDF/B-VIII library \cite{brown_endfb-viii_2018} for its contribution to the stopping of up-scattered fuel ions.

\subsection{Nuclear Fusion}

The fusion reactions we consider in our simulations are D--T, D--D, T--T and D--$^3$He using the parameterized cross sections from Bosch and Hale \cite{bosch_improved_1992}. We have also implemented the method presented in \cite{higginson_pairwise_2019} to capture the anisotropy of the D--T and D--D fusion reactions. For simplicity we treat D--$^3$He fusion as isotropic. For T--T fusion we use a Monte-Carlo implementation of the Lacina model \cite{lacina_neutron-neutron_1965}, which accounts for the neutron-neutron interaction in the $^4$He+n+n channel. We choose to neglect the two $^5$He+n channels as they are only a perturbative effect for the bulk $\sim$10 keV reactant energies in ICF \cite{casey_measurements_2012}. 

\section{ICF Neutron spectrum}

\subsection{D--T primary spectrum}

\begin{figure}[!htb]
    \centering
    \includegraphics[width=\linewidth]{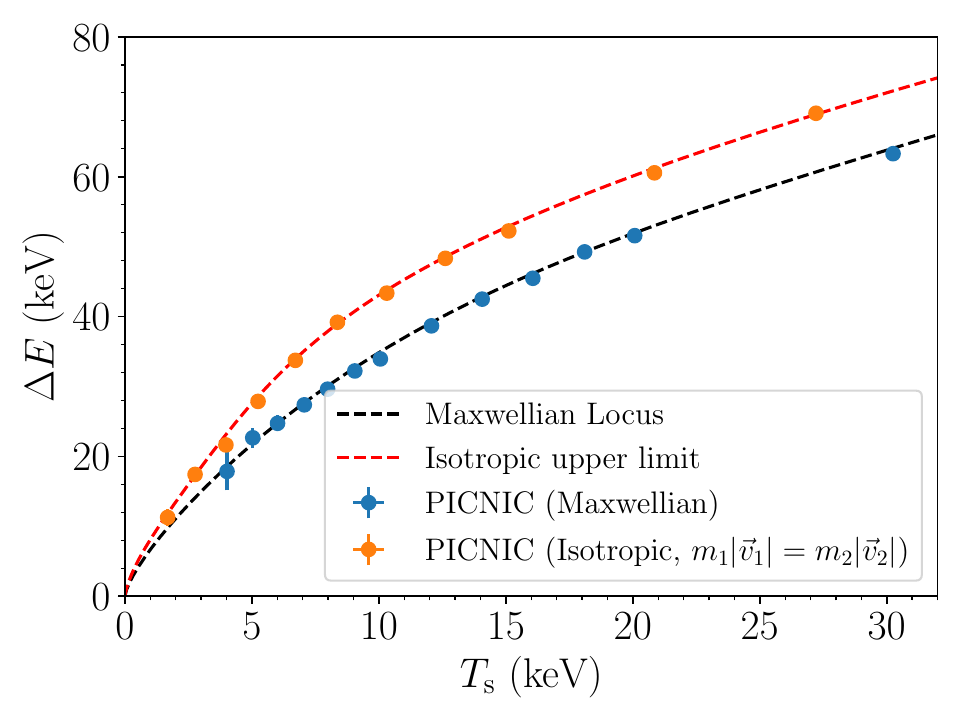} 
    \caption{Verification of the neutron spectra generated by the fusion algorithm implemented in PICNIC. The D--T neutron spectral shift and variance from simulations are compared with the Maxwellian locus and the isotropic distribution upper limit derived in \cite{ballabio_relativistic_1998} and \cite{crilly_constraints_2022} respectively. Simulations were performed on a 2D $10\times10$ grid with $N_\text{ppc}^\text{D} = N_\text{ppc}^\text{T} = 16384$ and were run for a single timestep to prevent time-variation of the particle distributions. The spectral temperature $T_\mathrm{s}$ was recovered using the sample variance of the neutrons. Particle motion and forces were turned off, making each simulation an average of 100 independent 0D (velocity space only) realizations.}
    \label{fig:MaxwellianLocus}
\end{figure}

The shift and variance of the spectral D--T neutron peak are measured to infer the yield-averaged ion temperature and relative kinetic energy between the reactants. The spectral shift is measured relative to the neutron energy at zero reaction energy $E_0 = 14.0284\text{ MeV}$
\begin{align}
	&\Delta E = \langle E_\text{n}\rangle - E_0\,,
\end{align}
where $\langle...\rangle$ denotes averaging over the total D--T neutron yield. The variance is directly related to the ion temperature via Doppler broadening \cite{brysk_fusion_1973}, with the spectral ion temperature defined as \cite{crilly_constraints_2022}
\begin{align}
	&T_\text{s} = \frac{m_\text{D}+m_\text{T}}{3\beta_0}\text{Var}\left(E_\text{n}\right)\,,
\end{align}
where $\beta_0 = 8852.7\text{ MeV}^2/c^2$. Given the distribution functions of the fuel ions, and assuming a steady state, there is a direct relation between the spectral shift and variance. For the special case of Maxwellian distribution functions, the relativistically correct relation between $\Delta E$ and $T_\text{s}$ was derived by Ballabio \cite{ballabio_relativistic_1998} and is shown in Fig.\,\ref{fig:MaxwellianLocus}, where it is compared with PICNIC simulation results recovering the same curve. This curve is often referred to as the ``hydrodynamic limit'', as any points that fall below it are accessible by Maxwellian plasmas with varying bulk fluid velocity Doppler shifts which increase the spectral variance. Spectra found above this curve, as reported in \cite{hartouni_evidence_2023}, would then be evidence of kinetic effects producing a specific class of non-Maxwellian distributions that produce larger spectral shifts. We also show in the same figure that PICNIC reproduces the upper limit curve accessible by isotropic distributions derived in \cite{crilly_constraints_2022}, which indicates the largest possible spectral shift accessible by isotropic velocity distributions.

\subsection{Reaction-in-flight neutrons}

RIF neutrons are produced in a 3-step process \cite{azechi_review_1991}. First, D--T fusion occurs between thermal ions, producing on average 3.54 MeV alphas and 14.1 MeV neutrons. Then, either the alphas or neutrons can ``knock-on'' D or T fuel ions by undergoing a large-angle elastic scatter. This generates a suprathermal D/T population which can then undergo fusion reactions with the thermal D/T in-flight, producing higher energy RIF neutrons. The kinematic limits of alpha knock-on and neutron knock-on (NKN) neutrons are summarized as follows \cite{jeet_diagnosing_2024}
\begin{align}
	&\alpha(3.5\!\text{ MeV})\!\rightarrow\!\text{T}(\leq\!3.4\!\text{ MeV})\!+\!\text{D}\!\rightarrow\!\text{n}(\text{10.6--20.6\! MeV})\,, \nonumber \\
	&\alpha(3.5\!\text{ MeV})\!\rightarrow\!\text{D}(\leq\!3.1\!\text{ MeV})\!+\!\text{T}\!\rightarrow\!\text{n}(\text{11.9--19.7\! MeV})\,, \nonumber \\
	&\text{n}(14.1\!\text{ MeV})\!\rightarrow\!\text{T}(\leq\!10.5\!\text{ MeV})\!+\!\text{D}\!\rightarrow\!\text{n}(\text{9.3--28.1\! MeV})\,, \nonumber \\
	&\text{n}(14.1\!\text{ MeV})\!\rightarrow\!\text{D}(\leq\!12.4\!\text{ MeV})\!+\!\text{T}\!\rightarrow\!\text{n}(\text{12.1--30\! MeV})\,.
\end{align}
The AKN part of the spectrum is the dominant RIF signal in the 15.5--18 MeV range, where the contributions from the Doppler-broadened D--T peak and NKN are smaller. This has led the AKN signal to be proposed as a direct probe for the alpha stopping power within the fuel during the burn \cite{hayes_reaction--flight_2015}. 

\subsection{Triton burn-up neutrons}

D--D fusion occurs via two equally likely branches whose products can lead to secondary fusion reactions
\begin{align}
	&\text{D}+\text{D} \rightarrow \text{T}(1.01\text{ MeV})\, + \text{p}(3.02\text{ MeV})\,,\nonumber\\
	&\text{D}+\text{D} \rightarrow \text{$^3$He}(0.82\text{ MeV})\, + \text{n}(2.45\text{ MeV})\,.
\end{align}
Aside from directly introducing a spectral peak at 2.45 MeV via the $^3$He+n channel, the T+p channel adds to the tails of the primary D--T spectrum with the emitted 1.01 MeV tritons. As these high energy tritons react with thermal deuterons, they produce ``triton burn-up'' (TBN) neutrons that appear in the neutron spectrum in the following energy range (in the absence of thermal broadening)
\begin{align}
	&\text{T}(\leq 1.01\text{ MeV}) + \text{D} \rightarrow \text{n}(\text{11.8 -- 17.1 MeV})\,.
\end{align}
Note that while this is the dominant non-Maxwellian signal for low-yield shots \cite{azechi_review_1991}, for high-yield ignition shots the TBN signal is drowned out by the RIFs by roughly an order of magnitude.

\subsection{T--T spectrum}

The kinematics of tritium-tritium fusion is complicated by its release of three products, allowing for a spectrum of possible product energies in the center-of-momentum frame. The T--T fusion reaction can follow three separate channels
\begin{align}
	&\text{T}+\text{T} \rightarrow \text{$^4$He} + \text{n} + \text{n}\,,\,\,\,\, Q=11.3\text{ MeV}\,, \nonumber \\
	&\text{T}+\text{T} \rightarrow \text{$^5$He} + \text{n}\,,\,\,\,\, Q_1=10.4\text{ MeV} \rightarrow \text{$^4$He} + \text{n} + \text{n}\,, \nonumber \\
	&\text{T}+\text{T} \rightarrow \text{$^5$He}^\ast + \text{n}\,,\,\,\,\, Q_1=9.2\text{ MeV} \rightarrow \text{$^4$He} + \text{n} + \text{n}\,. 
\end{align}
The kinematics for the first channel is well-described by the Lacina model \cite{lacina_neutron-neutron_1965}, which accounts for the neutron-neutron interaction. This skews what would otherwise be an elliptical 0--9.4 MeV T--T neutron spectrum towards lower energies and likewise skews the emitted alpha particle towards higher energies up to a maximum of 3.77 MeV. In this work we have opted to neglect the two $^5$He+n channels as their contribution is small at the $\sim$10 keV bulk reactant energies in ICF, only becoming important at energies above 100 keV \cite{casey_measurements_2012}. The first order correction to the neutron spectrum would be the addition of a small peak at 8.7 MeV from the ground state $^5$He+n channel.

\subsection{Fusion Anisotropy}

\begin{figure}[!htb]
    \centering
    \includegraphics[width=\linewidth]{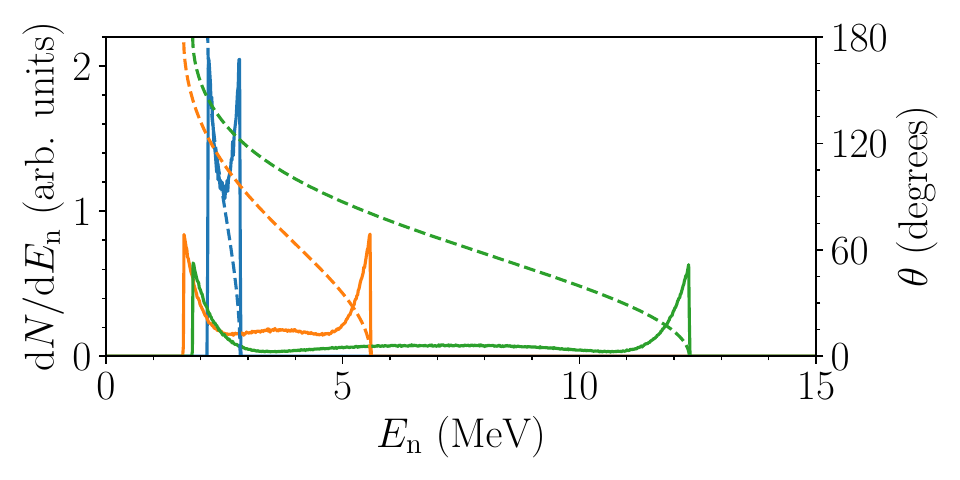} 
    \includegraphics[width=\linewidth]{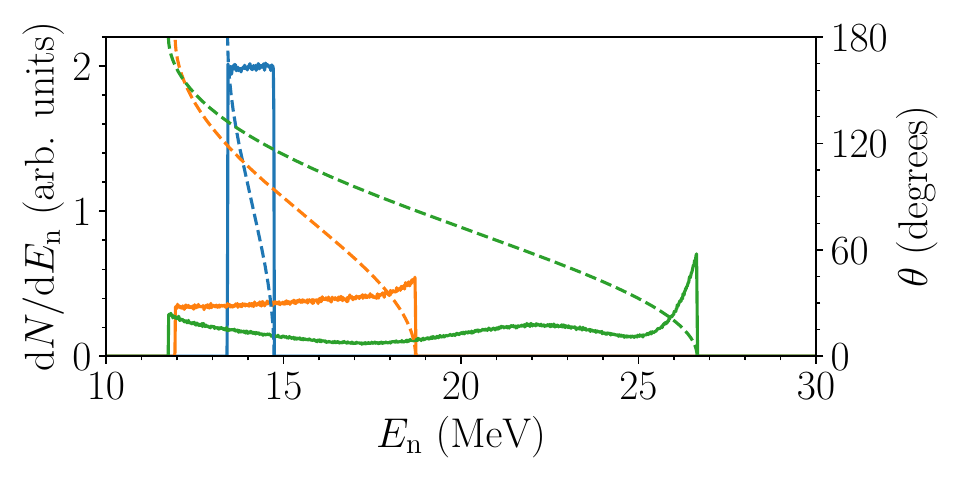} 
    \caption{Anisotropic neutron spectra (solid) and emission angle (dashed) from 0D cold beam-target fusion simulations for (top) D--D and (bottom) D--T. The deuteron beam proper velocities (blue, orange, green) are $u/c=1\%$, $5\%$ and $10\%$, which correspond to projectile energies of $\SI{93}{keV}$, $\SI{2.3}{MeV}$ and $\SI{9.3}{MeV}$ respectively.}
    \label{fig:anisoFusion}
\end{figure}

The RIF neutron spectrum is also affected by the anisotropy of D--T fusion at high energies. For reactant energies above $\sim$1 MeV, the emitted neutron becomes increasingly forward-biased with the incident deuteron in the center-of-momentum frame. This results in up-scattered deuterons reacting with a thermal triton having an increased likelihood of emitting upshifted neutrons, while up-scattered tritons reacting with a thermal deuteron tend to emit downshifted neutrons. Hence the high energy end of the AKN signal and the NKN signal will receive a larger contribution from the up-scattered deuterons and a reduced contribution from the up-scattered tritons. The neutron emission anisotropy is demonstrated in Fig.\,\ref{fig:anisoFusion}, which displays the lab frame neutron spectra of D--D and D--T beam-target fusion from PICNIC simulations with the implementation described in \cite{higginson_pairwise_2019} using the 2015 IAEA evaluation of the differential cross section \cite{otuka_evaluation_2015}.

\section{1D Spherical Simulation Results of Shot N210808}

We present the results of three simulations in 1D spherical geometry with the following scattering physics
\begin{enumerate}
\renewcommand{\labelenumi}{(\roman{enumi})}
\item cumulative Coulomb + isotropic fusion,
\item cumulative Coulomb + large-angle Rutherford + $\alpha$--D/$\alpha$--T NES + isotropic fusion,
\item cumulative Coulomb + large-angle Rutherford + $\alpha$--D/$\alpha$--T/D--T NES + anisotropic D--D and D--T fusion.
\end{enumerate}
All three simulations have bremsstrahlung and inverse bremsstrahlung and are initialized at approximately $\SI{150}{ps}$ before bangtime using density, temperature and implosion velocity profiles from 1D spherical HYDRA \cite{kritcher_design_2022,kritcher_design_2024} simulations of shot N210808 including the hotspot, D/T ice and carbon liner. This time was chosen to be sufficiently early before significant alpha heating occurs, while also already being hot enough to avoid having to account for degeneracy effects which can reduce the alpha stopping power \cite{maynard_born_1985}. We also assume that the D/T plasma at this stage is fully ionized in both the hotspot and the ice layer. 

\begin{figure}[!htb]
    \centering
    \includegraphics[width=\linewidth]{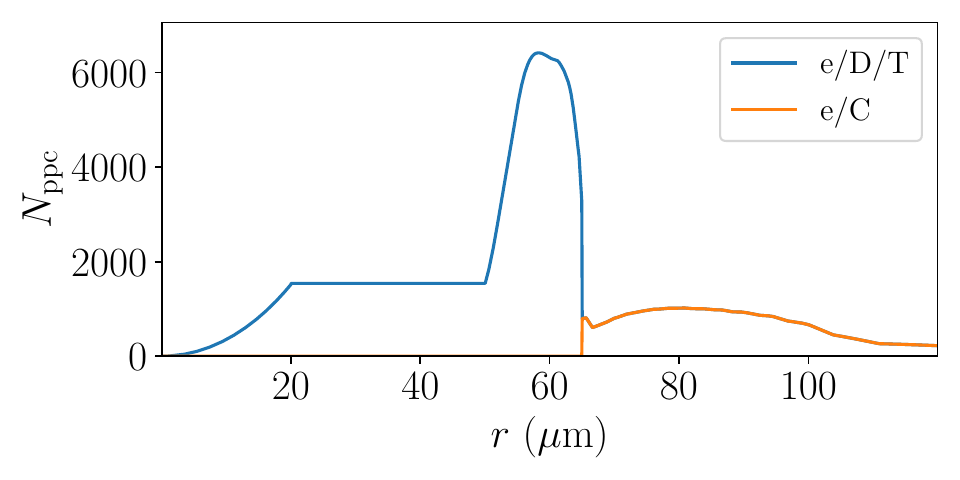} 
    \includegraphics[width=\linewidth]{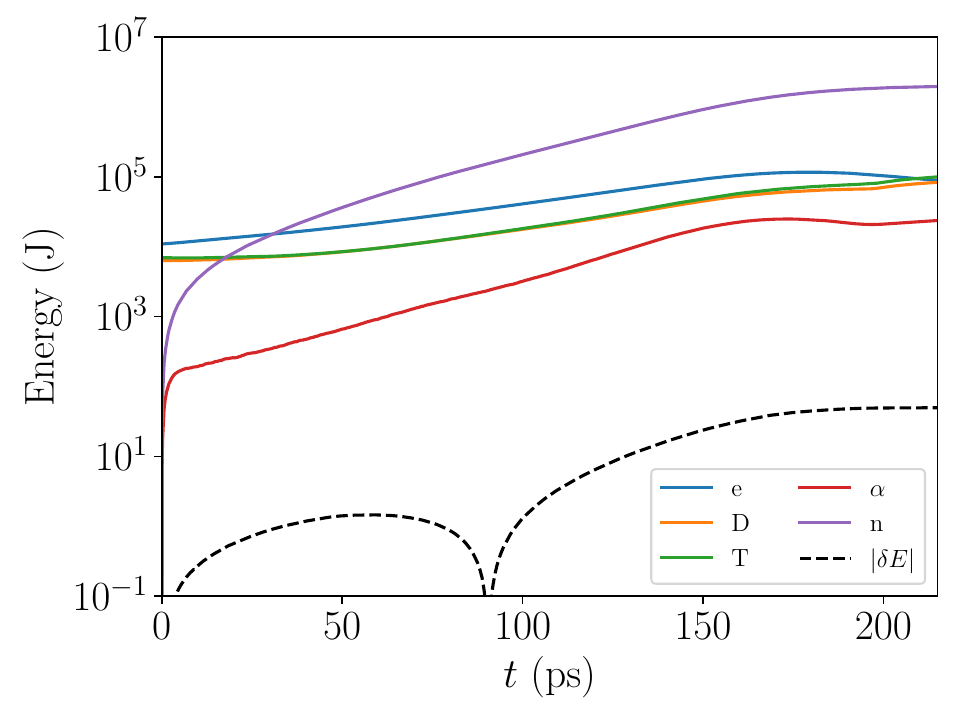} 
    \caption{(Top) number of macroparticles per cell of each species at initialization. (Bottom) total kinetic energy of each major species and the numerical energy conservation violation $\delta E$ of simulation (ii), which is representative of all three simulations.}
    \label{fig:Econs}
\end{figure}

As for the numerical setup, we use a uniform radial grid resolution of $\Delta r = \SI{120}{\micro m}/1728 = \SI{69.4}{nm}$ with a timestep of $\Delta t = \SI{0.1}{fs}$. We found this timestep sufficient in resolving the relevant collisional physics \cite{angus_binary_2025}. The number of particles per cell profiles of each species at initialization is given in Fig.\,\ref{fig:Econs}, with the total number of e/D/T/$^{12}$C macroparticles at initialization being $7.4\times10^6$. The numerical energy conservation is shown in the same figure, showing that the total energy conservation violation is several orders of magnitude smaller than the kinetic energies of the relevant species. We also use a simple particle splitting routine, which periodically checks for macroparticles with weights $4\times$ larger than the average in the cell (for cells with at least 20 macroparticles present) and splits them such that the new particle weights approximately match the cell average. We found this to be sufficient to prevent the comparatively mobile electrons from higher-weight regions from forming a non-physical sheath build-up at the origin, which inherently has low particle statistics in 1D spherical geometry. We did not find it necessary to use a particle merging algorithm for these simulations.

Neutron scattering is neglected in an effort to isolate the AKN signal and to avoid downscattered neutron noise in the low energy part of the spectrum, allowing us to also observe its possible effects on the D--D and T--T at-birth neutron spectra (as the fusion cross sections of D--T, D--D and T--T are all comparable in the $\sim$1 MeV range). We note that doing so forgoes both NKN and heating effects from neutrons depositing a portion of their energy in the plasma \cite{daughton_influence_2023,albright_use_2024}. 

\begin{figure}[!htb]
    \centering
    \includegraphics[width=\linewidth]{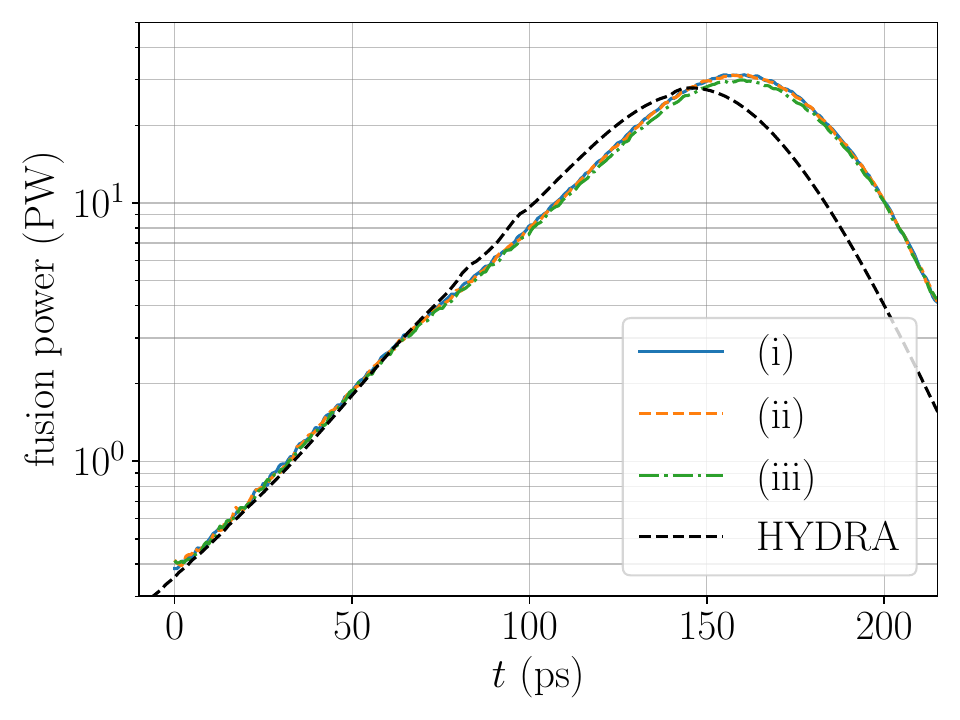} 
    \includegraphics[width=\linewidth]{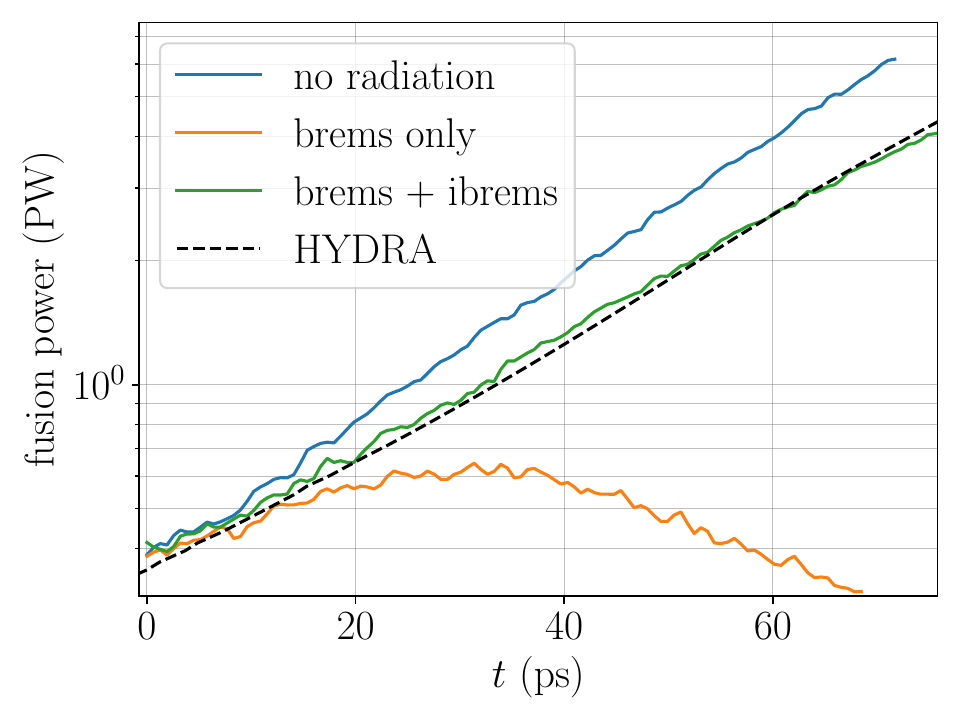} 
    \caption{(Top) fusion energy production rate comparison between PICNIC and HYDRA. (Bottom) early-time production rates of reruns of simulation (ii), one with no radiation physics and the other with bremsstrahlung emission, but no absorption.}
    \label{fig:yields}
\end{figure}
\begin{figure}[!htb]
    \centering
    \includegraphics[width=\linewidth]{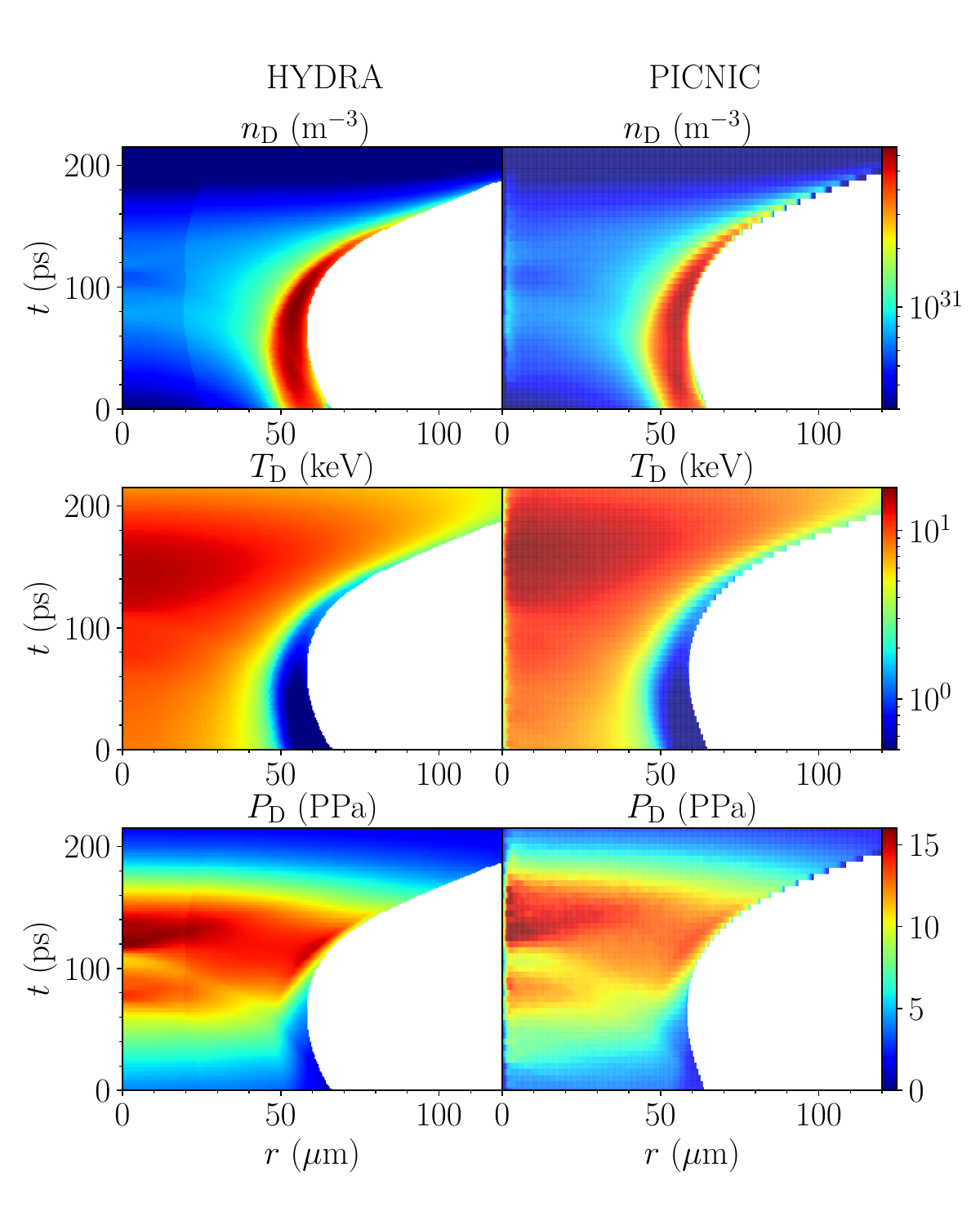} 
    \caption{Comparison of deuterium density, temperature and pressure phase-space plots between PICNIC simulation (i) and HYDRA.}
    \label{fig:n_T_comp}
\end{figure}

A comparison of the fusion production rates with HYDRA is shown in Fig.\,\ref{fig:yields}. Between the PIC simulations, we do not see any measurable differences in the yields with or without the inclusion of large-angle scattering physics in this regime, and only arguably see a slight reduction in yield with the inclusion of anisotropic fusion. Initially the PIC simulations have the same production rate as HYDRA. However, once alpha heating becomes relevant at around stagnation, the PIC fusion production rates begin to differ from HYDRA, resulting in a delayed bangtime and slightly increased yield. Determining whether this difference is a kinetic signature is subject for future work, as it requires a detailed comparison of the radiation, alpha stopping, and other forms of heat transport between PICNIC and HYDRA. To highlight the sensitivity to both the radiation emission and absorption models, we also include fusion production rate curves in Fig.\,\ref{fig:yields} of two extra runs of simulation (ii), one without radiation physics, which burns too hot, and the other with radiation but no absorption, which quickly quenches the reaction.

We compare the time evolution of the density, temperature and pressure profiles between PICNIC and HYDRA in Fig.\,\ref{fig:n_T_comp}, which qualitatively demonstrates that PICNIC can accurately capture 1D spherical hydrodynamics and burn propagation and similarly captures the anisobaric dynamics of the burn wave. The main discrepancy with HYDRA is delayed bangtime and a $\sim$20\% higher peak hotspot temperature, which is associated with the aforementioned increased yield prediction.

\begin{figure*}[!htb]
    \centering
    \includegraphics[width=\textwidth]{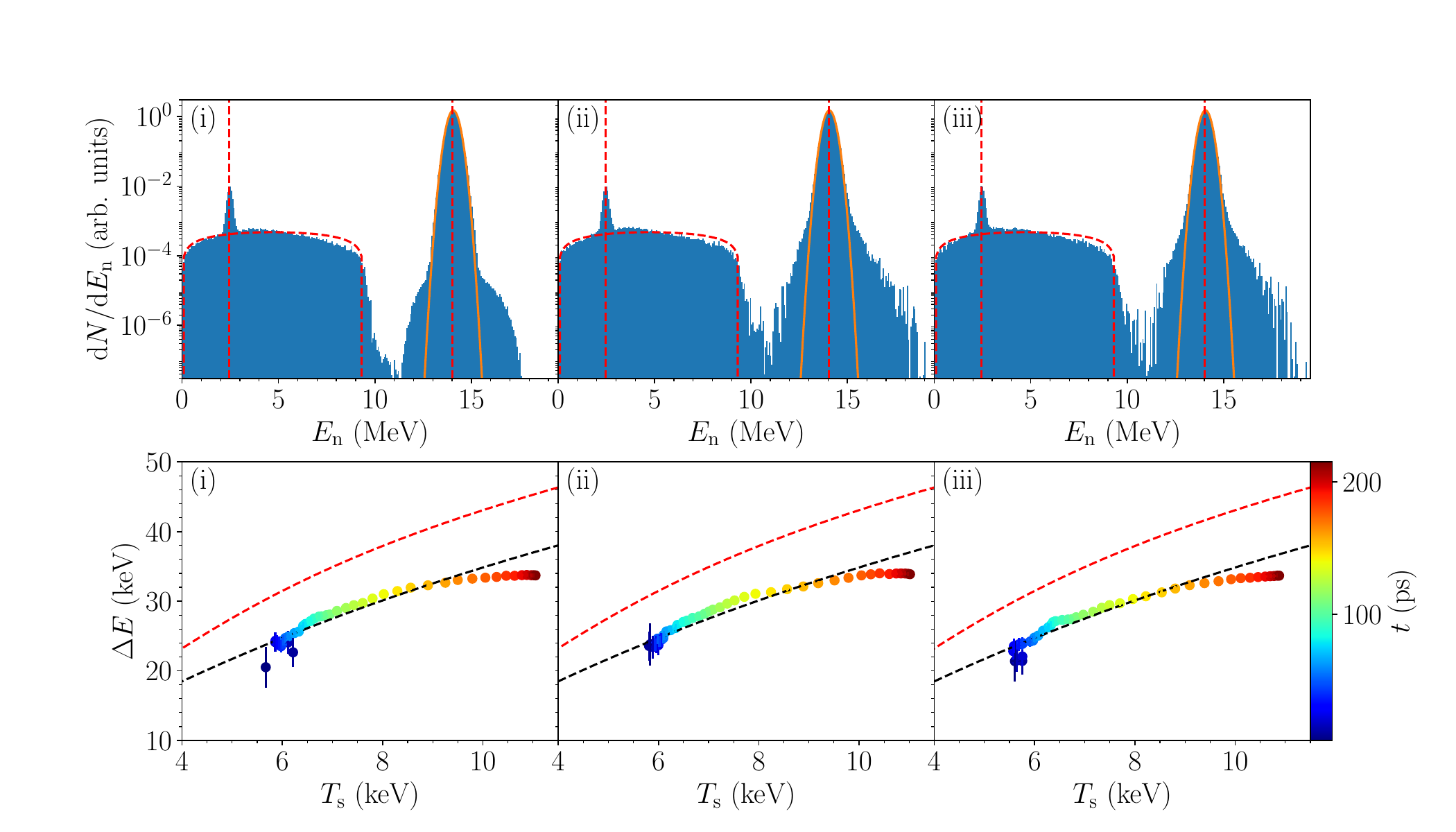} 
    \caption{(Top) total accumulated at-birth neutron spectra. The vertical red dashed lines indicate the D--D and D--T fusion neutron energies in the zero reactant energy limit $E_{0,\text{D--D}}=\SI{2.4487}{MeV}$, $E_{0,\text{D--T}}=\SI{14.0284}{MeV}$ \cite{crilly_constraints_2022}. Similarly, the red dashed region indicates the T--T fusion neutron spectrum at zero reactant energy without accounting for the neutron-neutron interaction. The solid orange curve is a fitted Gaussian to the D--T primary peak using the FWHM to set the variance, i.e. $\text{Var}(E_\text{n})=\text{FWHM}/(8\ln2)$. (Bottom) spectral shift vs. spectral temperature of the accumulated D--T primary neutron spectrum over time. The black and red dashed lines indicate the Maxwellian locus and the isotropic upper limit respectively. Error bars indicate the uncertainty due to low neutron macroparticle statistics at early times. The spectral temperature here is also inferred from the FWHM of the D--T primary peak.}
    \label{fig:nSpec}
\end{figure*}

The resulting neutron spectra are shown in Fig.\,\ref{fig:nSpec}. For the shallow-angle cumulative Coulomb only case (i), we can see the TBN contribution to the high energy neutron distribution, which is small relative to the AKN RIFs seen in cases (ii) and (iii). We do not see a notable difference in the D--D and T--T spectra apart from an increased broadening at the high end of the T--T spectrum due to RIFs. The 15.5--18 MeV AKN spectra from simulations (i) and (ii) are similar to those found in experiments at the NIF \cite{jeet_diagnosing_2024}. Above 18 MeV, the NKN signal in experiment begins to dominate, which was not included in our simulations. Markedly, there are no significant differences in the AKN spectra between cases (ii) and (iii), which is expected as D--T fusion is only slightly anisotropic in this reactant energy range as shown in Fig.\,\ref{fig:anisoFusion}. The D--T anisotropy is only envisioned to influence the high energy 18--30 MeV NKN signal in experiments. 

Shifting our attention to the spectral shift vs. temperature plots, at early times around stagnation the spectral shift hovers slightly above the Maxwellian locus. This suggests that non-Maxwellian distributions, apart from producing knock-on neutrons, also play a role in the burn itself. However, at bangtime and onwards the Doppler shifts from the hydrodynamic explosion widen the accumulated neutron spectrum such that it ends up well below the curve. Notably, we do not see this behavior change with the inclusion of large-angle scattering physics nor anisotropic fusion, and at no point in the simulations do the spectral shifts approach the anomalous 50--60 keV shifts measured experimentally in \cite{hartouni_evidence_2023}.

\begin{figure*}[!htb]
    \centering
    \includegraphics[width=0.49\linewidth]{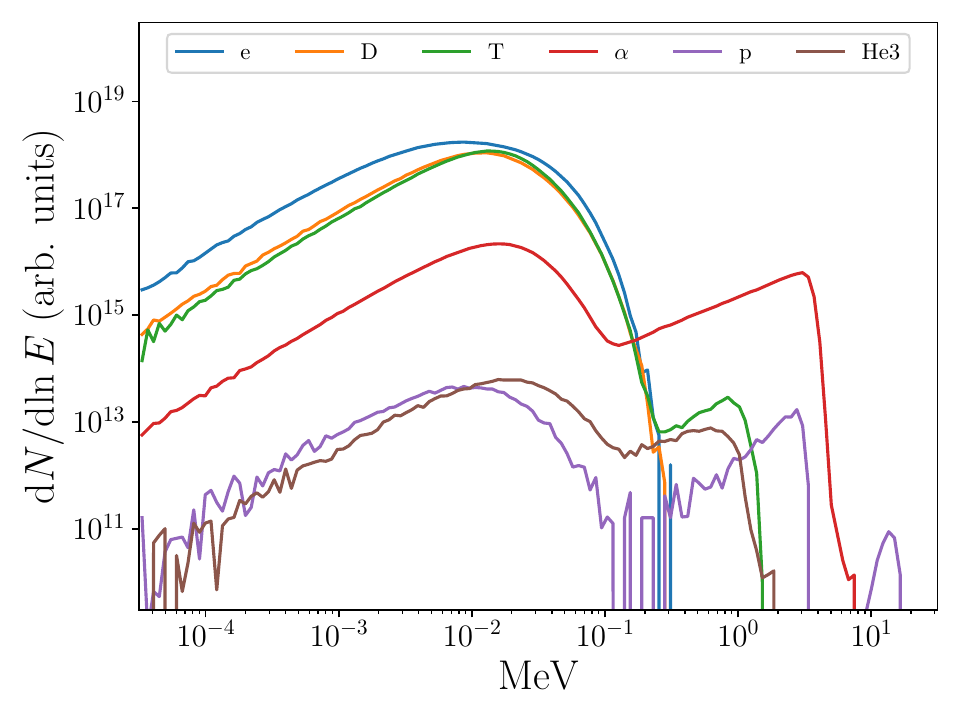} 
    \includegraphics[width=0.49\linewidth]{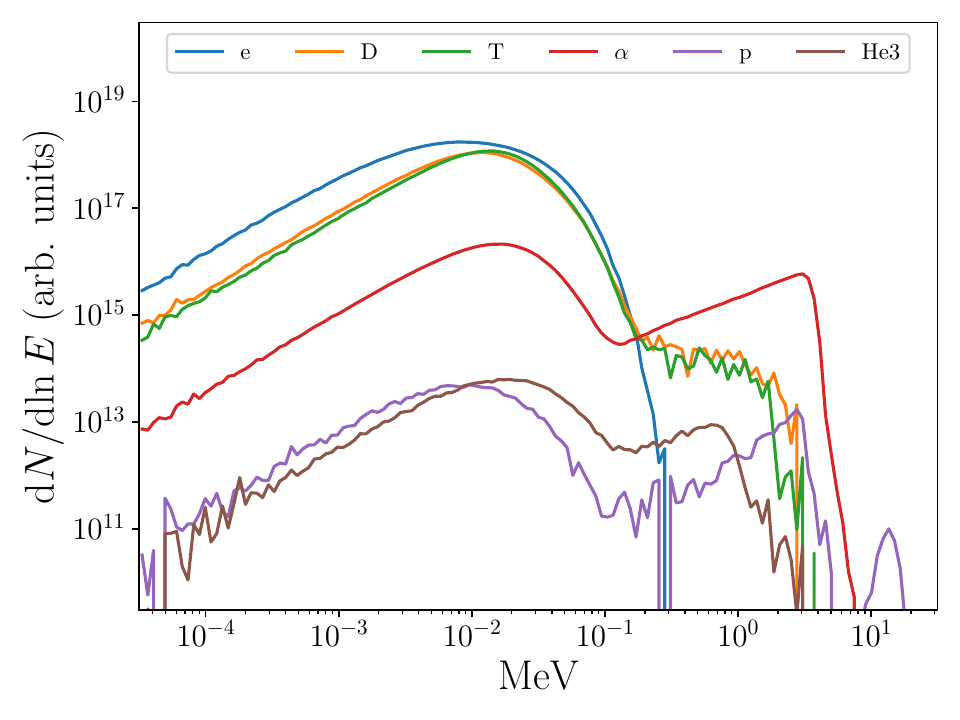} 
    \caption{Snapshots at bangtime of the energy spectra of all charged species in the fuel region for simulation (i) (left) with cumulative Coulomb scattering only and (ii) (right) with cumulative Coulomb, large-angle Rutherford and $\alpha$--D/$\alpha$--T NES.}
    \label{fig:charged_bt}
\end{figure*}

Snapshots of the energy spectra of the charged species at bangtime are shown in Fig.\,\ref{fig:charged_bt} to demonstrate the generation of suprathermal fuel ions via large-angle NES collisions with alphas, and its mediation by stopping against the bulk e/D/T plasma. The suprathermal D/T tail is similar to our previous 0D benchmarks for the generalized Coulomb method with large-angle Rutherford scattering \cite{angus_binary_2025}, but with a larger population relative to the alphas due to the inclusion of NES necessary to model AKN. Also present are the effects of large-angle Rutherford scattering on the minor species p and $^3$He, which receive small high energy tails above their D--D fusion birth energies. The small proton peak around 14.7 MeV is from D$^3$He fusion. For the shallow-angle cumulative Coulomb only case, there is still a source of suprathermal tritons from D--D fusion that produces TBN. \linebreak

\section{Conclusion}

We have presented results from 1D spherical simulations of the entire burnwave of the NIF shot N210808 using the code PICNIC, which treats all species fully kinetically with accurate masses, fields and scattering physics. This numerical tool allows us to study the kinetic effects involved in ignition, which are expected to become increasingly important with higher achieved yields, where the suprathermal fusion products become a significant fraction of the total mass.

With the inclusion of large-angle Rutherford and NES physics, we have recovered the well-known AKN RIF spectrum with this self-consistent, fully kinetic PIC code. However, despite having all the large-angle scattering physics necessary to produce the experimentally observed RIFs, we do not observe any large increase in the spectral shift of the D--T primary spectrum as was measured in \cite{hartouni_evidence_2023}. This rules out kinetic effects in 1D axisymmetric geometry with large-angle collisions as an explanation for the anomalous shift, which constrasts with the findings of a previous PIC study \cite{xue_mechanisms_2025}. Since 2D kinetic effects involving self-generated magnetic fields have not been ruled out, we plan to conduct simulations in RZ geometry in the future to investigate their impact on the burnwave and neutron spectrum. Kinetic effects associated with dopants or impurities in the fuel region, especially in the burning plasma regime where NES with alphas can generate suprathermal impurity ions, are also subject for future study.

\section*{Acknowledgments}
This work was performed under the auspices of the U.S. Department of Energy by Lawrence Livermore National Laboratory under Contract DE-AC52-07NA27344 and was supported by the LLNL-LDRD Program under Project No. 23-ERD-007. LLNL-JRNL-2006642.

\clearpage 

\bibliographystyle{model1-num-names}
\bibliography{refs}

\end{document}